\documentclass[runningheads]{llncs}
\usepackage{graphicx}
\usepackage{array}
\newcolumntype{P}[1]{>{\centering\arraybackslash}p{#1}}
\usepackage[caption=false]{subfig}
\usepackage{pgfplotstable}
\usepackage{pgfplots}
\begin{document}
\title{Supervised Initialization of LSTM Networks for Fundamental Frequency Detection in Noisy Speech Signals\thanks{Supported by the University of Costa Rica}}
%
%
\author{Marvin Coto-Jim\'enez\orcidID{0000-0002-6833-9938} }
\authorrunning{M. Coto-Jim\'enez}
%
\institute{PRIS-Lab, Escuela de Ingenier\'ia El\'ectrica.\\
Universidad de Costa Rica, San Jos\'e, Costa Rica.
\email{marvin.coto@ucr.ac.cr}}
\maketitle              
\begin{abstract}




Fundamental frequency is one of the most important parameters of human speech, of importance for the classification of accent, gender, speaking styles, speaker identification, age, among others. The proper detection of this parameter remains as an important challenge for severely degraded signals.
In previous references for detecting fundamental frequency in noisy speech using deep learning, the networks, such as Long Short-term Memory (LSTM) has been initialized with random weights, and then trained following a back-propagation through time algorithm. In this work, a proposal for a more efficient initialization, based on a supervised training using an Auto-associative network, is presented. This initialization is a better starting point for the detection of fundamental frequency in noisy speech. The advantages of this initialization are noticeable using objective measures for the accuracy of the detection and for the training of the networks, under the presence of additive white noise at different signal-to-noise levels. 

\end{abstract}
\section{Introduction}

The analysis of noisy speech signals has been a topic of interest over the past several decades. The main reason for this interest relies on the distortion of those signals in real-world environments. The quality of the communications systems or the recognition performance may be affected in their quality~\cite{weninger2014,weninger2014b,narayanan2013,bagchi2015} with such noise and distortion added to the speech information. 

The speech enhancement algorithms developed to reduce the noise or detect parameters in noisy speech can be viewed as successful if they reduce perceivable background noise, preserve or enhance signal quality~\cite{hansen1998} or provide a better detection of relevant parameters, such as fundamental frequency ($f_0$). 

Traditional signal processing-based methods, such as the Wiener filtering and Spectral Subtraction, among many others, provide noise reduction based on signal processing algorithms. More recently, Deep Neural Networks (DNN) have been presented in~\cite{du2014,han2015,maas2012,li2013}. The main approach for DNN is the mapping of spectral features from noisy speech into the features of the corresponding clean speech.

Among the new types of neural networks with recurrent connections (usually named Recurrent Neural Networks or RNN), the Long Short-Term Memory Network (LSTM) has succeeded over other types of networks in reducing noise and reverberant distortions in speech signals. The most common features used in those applications are the Mel-Frequency Cepstrum Coefficients (MFCC), derived from the spectrum information~\cite{mfcc}. 

One of the most important tasks in noisy speech processing is the $f_{0}$ detection. Accurately detecting this parameter is of importance in many applications. There are still demands for improvement in the algorithms proposed so far, especially for noisy speech~\cite{wu2016}.

In this work, motivated by the success of LSTM in speech enhancement, we present a novel way to initialize LSTM neural networks to enhance the $f_0$ detection in speech degraded with noise. The initialization of the network is based on a supervised Auto-associative network that learns the identity function between its inputs and its outputs and presents a better starting set of weights for the training process for the LSTM. With the LSTM initialized this way, we show the benefits for the $f_0$ detection in several Signal to Noise Ratio (SNR) levels.

\subsection{Related work}

Previous studies on robust $f_{0}$ detection with signal processing-based techniques have explored the harmonic information of the signals in the frequency domain or periodicity in the time domain~\cite{han2014}. More recently, deep learning algorithms, consisting of several layers of neural networks, have been used in several works related to noise reduction and parameter detection of noisy speech, especially the enhancement of features derived from the spectrum, typically MFCC~\cite{seltzer2013investigation,abdel2012applying,huang2013audio}.

The principal mechanism for enhancing speech signals using deep learning algorithms is to apply DNN as regression models, mapping the noisy speech parameters into the corresponding clean speech parameters~\cite{weninger2014}\cite{weninger2014b}. 

Some of these applications of deep learning have outperformed other denoising algorithms on speech signals containing noise of different types with various signal-to-noise levels~\cite{kumar2016,hinton2012deep,coto2017}. Reverberant speech has also been analyzed with this approach~\cite{feng2014speech,ishii2013}.

For example, in~\cite{nakashika-2013}, the spectrum features of the synthesized speech are enhanced using models such as DBNs, and also RBM has been previously studied~\cite{chen-2015}. More recently, the use of Recurrent Neural Networks (RNNs) was presented, with the advantage of the inherent structure of RNNs, which seems to deal better with the time-dependent nature of the speech signal paramters. 

In these references, the most common approach is to enhance the spectral components of the synthetic speech, by mapping them to those of the original ones, using diverse deep learning algorithms. A more recent approach that considers a single-stage of multi-stream post-filters have been presented in~\cite{coto-2017}\cite{coto-2016c}, with greater success than the single post-filters based on LSTM.

In $f_0$ detection, LSTM networks have recently outperformed several other algorithms~\cite{liu2017}. We use this previous work and provide a better initialization state for the task. The training process of neural networks, such as those presented in the following subsection, is traditionally based on a random initialization of weights that represents connections between inputs, hidden units or outputs. Following the initialization, a process of adjustment of those weights based on the data presented to the network is performed, using algorithms such as back-propagation or back-propagation through time, depending on the type of neural network.  

To increase the effectiveness of the DNN, unsupervised initialization and then fine-tuning processes with other networks, such as Restricted Boltzmann Machines (RBM), have been presented~\cite{dahl2012}. It is commonly considered the breakthrough of effective training for deep neural networks the algorithms for training deep belief networks (DBN), based on a layer-wise unsupervised pre-training followed by supervised fine-tuning~\cite{erhan2010}. 

The benefits of unsupervised pre-training stages before the training algorithms have been also verified in fields other than speech processing, such as music classification~\cite{vanden2014} and visual recognition~\cite{donahue2014}. Those unsupervised approaches present data at the input of the neural network and affect the weights without comparing the output to the corresponding data. Semi-supervised techniques have been also applied in similar applications~\cite{vesely2013} combining at least one stage of unlabeled data to initialize the weights of the neural networks. 

In our approach, the initialization is completely supervised, with an Auto-associative network trained to map the identity function between its inputs and its outputs, in which weights become the initial weights of the LSTM network. In $f_{0}$ detection, this Auto-associative network provides better results in signals degraded with white noise, due to the relationship established between the information provided in the initial stated of the network.

\subsection{Problem Statement}

In noise reduction for speech applications, a speech signal degraded with additive noise is processed to improve its quality or detect parameters distorted by the noise. We can assume that the corrupted signal, $y$, is the sum of a clean speech signal, $x$, and noise $d$, i.e.:

\begin{equation}
y(t)=x(t)+d(t)
\end{equation}




In deep learning-based approaches, $x(t)$ can be estimated directly from data, using algorithms that learn a mapping function $f(\cdot)$ between noisy and clean data:

\begin{equation}
\hat{x}(t)=f\left(y(t)\right).
\end{equation}

The precision of the approximation $f(\cdot)$ depends on factors such as the amount of training data, the kind of neural network, its architecture and the algorithm selected.

During the training of the network, the initial set of weights of the network $\theta_{i}$ is commonly established as random numbers, and are updated during the training process until stop criteria, defined as a maximum number of epochs or minimum validation error. At the end of the process, a set of updated and probably completely different weights $\theta_{f}$ are stored and used with a test set. 

With the supervised initialization proposed in this work, the mapping of identity function from the inputs to the outputs is pre-trained, providing a set of parameters $\theta_{A}$ that are close of those of $\theta_{f}$, due to the denoising nature of the regression problem and $f_{0}$ detection of similar data.

In terms of broad sound classification $f_{0}$ has two important values: Voiced and Unvoiced. Voiced sounds present values of $f_{0}>0$, and are present on vowels and some consonants, and Unvoiced sounds are present in silence and many consonants, such as /b/, /p/ and /s/. 

We pretend to show that $\theta_{A}$ is a better initialization for the LSTM than the random $\theta_{R}$, in terms of detection rate and voiced/unvoiced decision error. 

To our knowledge, this is a novel way to initialize and employ LSTM networks to detect $f_0$ in noisy speech. The rest of this article is organized as follows: Section 2 provides some details of the LSTM neural networks. Section 3 describes the proposed systems and the experiments carried out to test the proposal. Section 4 presents and discuss the results, and finally, conclusions are given in Section 5.

\section{Long Short-term Memory Neural Networks}

In speech enhancement of noisy speech and detection of $f_{0}$ under noisy conditions, several groups of researchers have experimented with deep learning algorithms. Recurrent Neural Networks (RNN)~\cite{ocho}, which includes feedback from neurons to themselves and others on the same layer, have achieved particularly good results, particularly in modeling the dependent nature of speech parameters. LSTM have been presented in~\cite{trece} as an extended RNN, with the capacity of learn long term relationships in data, and store information for long or short time intervals. 

Among the many successful implementations of LSTM are automatic speech recognition systems, speech synthesis and handwriting generation, where past values of the parameters are important to classify of perform regression~\cite{diez}~\cite{once}. 

LSTM has a structure similar to those of basic RNN: a set of units inputs the sequences $\mathbf{y}=\left(y_{1},y_{2},\dots,y_{T} \right)$, and hidden vector sequences $\mathbf{h}=\left(h_{1},h_{2},\dots,h_{T} \right)$ are calculated through the set of weights between inputs and hidden units, of between hidden and hidden units of the next layer. 



Detailed mathematical description of the LSTM networks can be found on~\cite{ocho}\cite{trece}\cite{gers2002}. In this work we have followed the implementation described in~\cite{coto2017}.

\subsection{$f_0$ detection with autoencoders}

In deep learning approaches for $f_0$ detection, a neural network is trained with several inputs, from where the $f_{0}$ can be inferred. The parameters of the network are found using training data to minimize the average reconstruction of the input, that is, to have output $f(y)$ as close as possible to the uncorrupted signal $x$~\cite{vincent2010stacked}, particularly on the $f_{0}$ parameter.

One of the recent architectures of neural networks applied in enhancing noisy speech is the denoising autoencoder, which consists of two parts: the first part is the encoder, where a mapping $f$ transforms an input vector $y$ into a representation $h$ in the hidden layers. The second part is the decoder, where a mapping from the hidden representation into a vector $\hat{x}$ is performed. 

For the purpose of $f_{0}$ detection in noisy speech, during the training stage, noise-corrupted parameters are presented at the inputs of the autoencoders, while the corresponding clean features of the same dimensionality became the outputs. The training algorithm adjusts the parameters of the network in order to learn the complex relationships between them, and output $f_0$ which corresponds to the detected parameters from the noisy speech.

\section{Proposed system}
To detect $f_{0}$ from noisy speech, the mapping from noisy $f_{0}$ can be learned directly from the data~\cite{tres}, with training, validation and test sets and procedures traditionally defined for machine learning algorithms. For this purpose, we use sentences of noisy utterances and the corresponding clean version to train the LSTM autoencoder networks.


The weights of the LSTM networks are initialized in two ways:

\begin{itemize}
\item Randomly: All the weights have random numbers at the first epoch of training. This is the most common usage in this application and denoising-related tasks, and we take it as the base system. 
\item Auto-associative: An auto-associative network is a neural network whose input and target vectors are the same~\cite{baek2003}. Here, the LSTM networks are trained presenting the same clean data at the input and at the output in each frame. This way, the network learns the identity function between its inputs and its outputs. After training, the weights of the Auto-associative networks became the initialized weights of the corresponding LSTM networks for $f_0$ detection.

\end{itemize}


\subsection{Corpus description}

In our experiments, we use the CMU Arctic database, described in~\cite{cmu}. We chose one US-English voice: SLT. The dataset is phonetically balanced, originally designed for research in unit selection speech synthesis, and consist of around 1150 utterances selected from out-of-copyright texts from Project Gutenberg.

\subsection{Feature extraction}

The audio files of the database were downsampled to 16kHz, in order to extract the parameters using the Ahocoder system. In this system, the fundamental frequency $f_0^k$ (zero-valued if invoiced), 39 MFCC, plus an energy coefficient are extracted from each frame. Hence each frame is represented by a 41th dimensional vector $V_{k} = \left[ f_0^k, e^k , mfcc_k^1,\dots, mfcc_k^{39}\right]$. Details on the parameter extraction and waveform regeneration of the Ahocoder system can be found in~\cite{erro2}.

\subsection{Auto-associative initialization}
The initialization procedure was performed using 800 utterances of the clean SLT voice, with the same parameters at the input and at the output of the network. The usual forward and back-propagation through time algorithms were applied, and the stop criteria was 40 epochs from the last best result, or a maximum of 1000 epochs.

\section{Experimental setup}

We shall describe in some detail the experimental setup that was followed in this work. The whole process can be summarized in the following steps:

\begin{enumerate}

\item Noisy database generation: Files containing white noise were generated and added to each audio file in the database for five SNR, in order to cover a range from light to heavy noise levels. 

\item Feature extraction and input-output correspondence: A set of parameters was extracted from the noisy and the clean audio files. Those from the noisy files were used as inputs to the networks, while the corresponding clean features were the outputs. 

\item Training: During training, the weights of the networks were adjusted as inputs, and the clean features of the audio files were presented. A validation set of 150 sentences was also used.

\item Test: A subset of 50 randomly selected utterances was chosen for the test set. These utterances were not part of the training process, to provide independence between the training and testing. 
\end{enumerate}

In order to determine the improvement in the efficiency of the supervised pre-training, the following objective measures were adopted~\cite{liu2017}:

\begin{itemize}
\item DR (Detection Rate): Evaluated on voiced frames, where a $f_{0}$ estimate is considered correct if the 
deviation of the estimated $f0$ is within 5\% of the ground truth clean value of $f_{0}$.

\begin{equation}
DR=\frac{N_{0.05}}{N_{p}}\times 100\%,
\end{equation}

where $N$ represent number of frames.

\item VDE (Voice Decision Error): Indicates the percentage of frames misclassified in terms of voicing/unvoicing: 

\begin{equation}
VDE=\frac{N_{V\rightarrow U}+N_{U\rightarrow V}}{N}\times 100\%,
\end{equation}
 
 where $N_{V\rightarrow U}$ and $N_{V\rightarrow U}$ represent misclassification of Voiced or Unvoiced frames.
 
 \item Sum of squared erros (sse): This is a common measure for the error in the validation and test sets during training. It is defined as:

\begin{eqnarray}
\mbox{sse}(\theta) & = & \sum_{n=1}^{T}\left(\mathbf{c_{x}}-\hat{\mathbf{c_{x}}} \right)^{2}\\
                   & = & \sum_{n=1}^{T}\left(\mathbf{c_{x}}-f(\mathbf{c_{x}}) \right)^{2},
\end{eqnarray}
 
 where $c_{x}$ is the known value of the outputs and $\hat{c_{x}}$ the its approximation from the network.
 
\end{itemize}

\section{Results and Discussion}

We present the results of three systems considered in this work, related to the initialization of the networks and a established algorithm to detect $f_{0}$ in speech signals. To allow a comparison with the base Ahocoder $f_{0}$ detection algorithm, the $f_0$ values that follows correspond to $\log(f_{0})$. The three systems considered are:

\begin{enumerate}
\item None: $f_{0}$ detection directly from the noisy speech, provided with the algorithm implemented in the Ahocoder system, based on harmonic analysis.
\item LSTM: $f_{0}$ detection with the LSTM network initialized with random weights.
\item LSTM-AA: $f_{0}$ detection with the weights of the LSTM network initialized from the Auto-associative network.
\end{enumerate}

Table~\ref{tablelstma1} shows the results of the DR for the three systems and the five levels of white noise:

\begin{table*}[htbp]
\begin{center}
\caption{Comparison of the results for the VDE in the detection $f_{0}$. Lower values represent better results}
\label{tablelstma1}
\begin{tabular}{cP{2cm}P{2cm}P{2cm}}
 SNR & None & LSTM & LSTM-AA  \\
\hline
 -10 &  67.23\%  &  5.06\%    & 5.20\%        \\
 -5  & 58.79\% &  3.70\%    & 3.67\%        \\
 0  &  25.20\%    &  3.16\%    & 2.99\%        \\
 5  &   10.02\%   &  5.81\%   & 5.75\%        \\
 10  &  4.87\%   & 7.83\%     & 7.82\%       \\
\hline
\end{tabular}
\end{center}
\end{table*}

With the exception of SNR-10, the Auto-associative initialization presents better values of VDE in all SNR levels. The exception can be explained in terms of the particularly different parameters of the clean speech used at the initialization of the network, which are very different from those of the noisy signal at the inputs. The rest of results verified that the Auto-associative initialization allow the LSTM to provide better Voiced/Unvoiced decisions.  

\begin{figure*}[htbp]
\begin{center}


SNR -5
\begin{tikzpicture}
\begin{axis}[
  width=\textwidth,
  height=0.3\textwidth,
  grid=major, 
  grid style={dashed,gray!30},
  mark size=0.5pt,
  xlabel=epoch,
  ylabel=sse]
\addplot table [y=Random, x=Epoch]{snr-5.dat};
\addlegendentry{Random}
\addplot table [y=Auto-Associative, x=Epoch]{snr-5.dat};
\addlegendentry{Auto-Asoc.}
\end{axis}
\end{tikzpicture}

SNR 0
\begin{tikzpicture}
\begin{axis}[
  width=\textwidth,
  height=0.3\textwidth,
  grid=major, 
  grid style={dashed,gray!30},
  mark size=0.5pt,
  xlabel=epoch,
  ylabel=sse]
\addplot table [y=Random, x=Epoch]{snr0.dat};
\addlegendentry{Random}
\addplot table [y=Auto-Associative, x=Epoch]{snr0.dat};
\addlegendentry{Auto-Asoc.}
\end{axis}
\end{tikzpicture}

SNR5
\begin{tikzpicture}
\begin{axis}[
  width=\textwidth,
  height=0.3\textwidth,
  grid=major, 
  grid style={dashed,gray!30},
  mark size=0.5pt,
  xlabel=epoch,
  ylabel=sse]
\addplot table [y=Random, x=Epoch]{snr5.dat};
\addlegendentry{Random}
\addplot table [y=Auto-Associative, x=Epoch]{snr0.dat};
\addlegendentry{Auto-Asoc.}
\end{axis}
\end{tikzpicture}


\caption{Sample of evolution of sse error in each epoch for the LSTM network with random and Auto-associative initialization. The lower sse value means better results, while the fewer epochs it takes to reach the minimum sse represent a more efficient training.}
\label{figuresse}
\end{center}
\end{figure*}
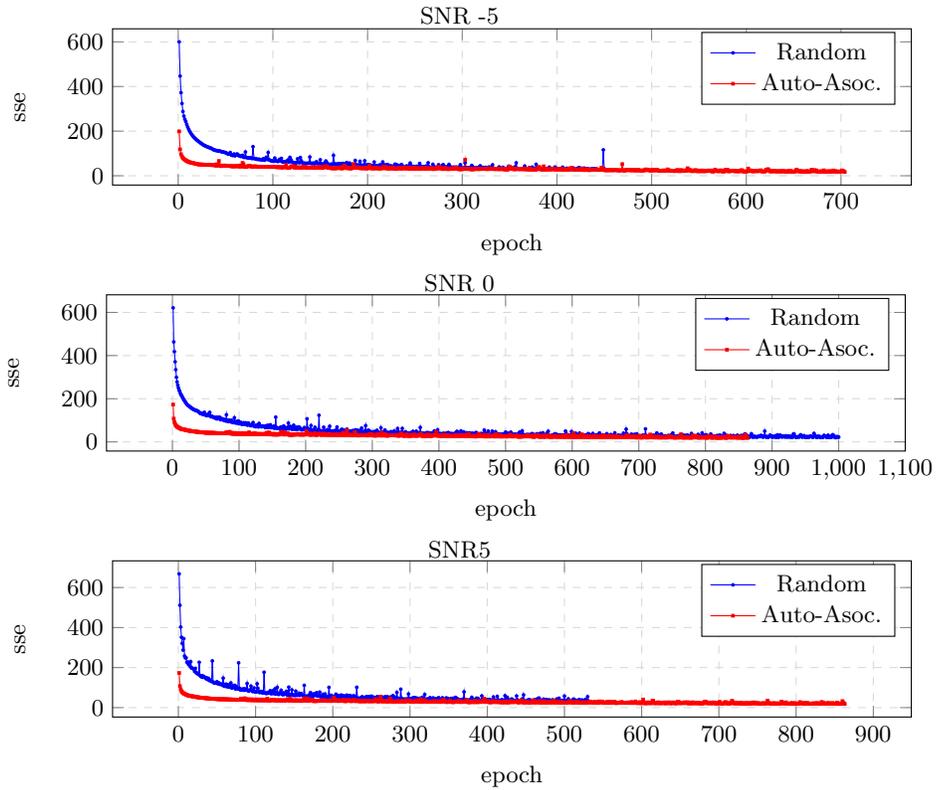

This better performance in VDE could benefit automatic speech recognition systems and perceptual quality of the signals. One additional advantage of the LSTM-AA is the more efficient training time of the network and lower sse error achieved. For example, Figure~\ref{figuresse} shows the evolution of the sse on the validation set during training.

SNR-5 and SNR5 were the only case where the LSTM-AA required more epochs to achieve the best set of parameters of the network. In these cases, the significant lower value of sse and the benefit in terms of the objective quality measures allow us to consider a positive result. In the five SNR levels, LSTM-AA presents lower values in comparison to LSTM in almost every epoch, and for the case of SNR0, SNR10 and SNR-10 the best value were reached in fewer epochs.

The results of the DR measure are shown in Table~\ref{tabla2}. Similar to the VDE measure, the Auto-associative initialization of the LSTM network present better results than the random initialization, with the exception of SNR-10. Significant decreases of DR value at SNR0 and SNR-5 are consistent with the better values obtained in the VDE measure at this levels.

\begin{table*}[htbp]
\begin{center}
\caption{Comparison of the results for the DR in the detection $f_{0}$ in voiced frames. Lower values represent better results.}
\label{tabla2}
\begin{tabular}{cP{2cm}P{2cm}P{2cm}}
 SNR & None & LSTM & LSTM-AA  \\
\hline
 -10 &  100\%    &   7.12\%    &  8.96\%        \\
 -5  &  88.85\%  &   7.86\%    &  4.52\%        \\
 0  &  43.84\%  &   5.30\%    &  4.21\%        \\
 5  &   15.31\%  &  2.94\%   &  2.43\%        \\
 10  &  7.36\%   &  1.06\%     &  1.24\%       \\
\hline
\end{tabular}
\end{center}
\end{table*}

This results shows how the initialization proposed benefit the $f_{0}$ detection also in terms of the precision of the $f_{0}$ detection.

\begin{figure*}[htbp]
\begin{center}
\includegraphics[width=\textwidth]{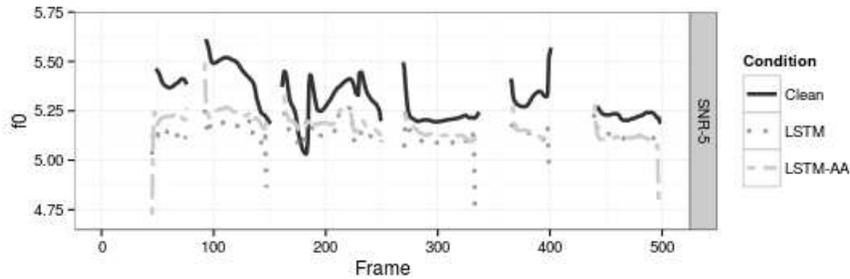}
\caption{Contours of $f_{0}$ for SNR-5.}
\label{figure2}
\end{center}
\end{figure*}

To provide a visual representation of the benefit of our proposal, Figure~\ref{figure2} shows $f_{0}$ contours for one utterance, and compare the contour after detection of the Ahocoder algorithm in noisy speech, $f_0$ detected with LSTM initialized with random weights, and the $f_{0}$ value detected with the LSTM initialized with the Auto-associative network.

It is confirmed in the $f_{0}$ contour that the values from LSTM-AA are closer to those of the natural speech. It is important to note that for this SNR level, the Ahocoder $f_{0}$ extraction has 58.79\% in VDE and 88.85\% in DR measures.

\section{Conclusions}

In this work, we have presented a proposal for initializing LSTM networks, with the objective of enhancing the $f_{0}$ detection in noisy speech. We conducted a comparison using five levels of White noise, two well-known measures for the $f_{0}$ detection and additional evaluation for training efficiency of the networks.

It is remarkable the capacity of LSTM networks to detect $f_{0}$ in heavy noisy conditions, where the $f_{0}$ is totally undetectable for other algorithms. Our proposal of Auto-associative initialization performed better in terms of a more efficient training time and sse of the LSTM, and lower values of Detection Rate of Voiced/Unvoiced frames and Voice Decision Error in comparison with the traditional random initialization.

The main assumption for applying the initialization was that Auto-associative network provides a better approximation for the regression performed from the noisy speech to clean parameters in the LSTM, with the exception of the lower SNR value.

Future work will include the application of this proposal for the enhancing of all parameters of the speech signal, so the benefits can be also tested in terms of other objective and subjective measurements for the enhancement of noisy speech signals. Also, the combination of signal processing-based denoising algorithms with LSTM networks should benefit the whole process of speech enhancing and $f_{0}$ detection.

%
%
%

\begin{thebibliography}{50}

\bibitem{weninger2014} Weninger, F., Watanabe, S., Tachioka, Y. and Schuller, B. ``Deep recurrent de-noising auto-encoder and blind de-reverberation for reverberated speech recognition''. In Acoustics, Speech and Signal Processing (ICASSP), 2014 IEEE International Conference on (pp. 4623-4627). IEEE.

\bibitem{weninger2014b} Weninger, F., et al. ``Feature enhancement by deep LSTM networks for ASR in reverberant multisource environments.'' Computer Speech \& Language 28.4 (2014): 888-902.

\bibitem{narayanan2013}
Narayanan, A., Wang, D. ``Ideal ratio mask estimation using deep neural networks for robust speech recognition.'' 2013 IEEE International Conference on Acoustics, Speech and Signal Processing. IEEE, 2013.

\bibitem{bagchi2015} Bagchi, D., Mandel, M.I., Wang, Z., He, Y., Plummer, A. and Fosler-Lussier, E. ``Combining spectral feature mapping and multi-channel model-based source separation for noise-robust automatic speech recognition.'' In Proc. IEEE ASRU 2015.

\bibitem{hansen1998} Hansen, J., Pellom, B.L. "An effective quality evaluation protocol for speech enhancement algorithms." ICSLP. Vol. 7. 1998.

\bibitem{du2014} Du, J., Wang, Q., Gao, T., Xu, Y., Dai, L.R. and Lee, C.H., 2014. Robust speech recognition with speech enhanced deep neural networks. In INTERSPEECH (pp. 616-620).

\bibitem{han2015} Han, K., He, Y., Bagchi, D., Fosler-Lussier, E. and Wang, D., 2015. Deep neural network based spectral feature mapping for robust speech recognition. In Proc. Interspeech (pp. 2484-2488).

\bibitem{maas2012} Maas, A.L., Le, Q.V., O'Neil, T.M., Vinyals, O., Nguyen, P. and Ng, A.Y., 2012, September. Recurrent Neural Networks for Noise Reduction in Robust ASR. In INTERSPEECH (pp. 22-25).

\bibitem{li2013} Deng, L., Li, J., Huang, J.T., Yao, K., Yu, D., Seide, F., Seltzer, M., Zweig, G., He, X., Williams, J. and Gong, Y., 2013, May. Recent advances in deep learning for speech research at Microsoft. In Acoustics, Speech and Signal Processing (ICASSP), 2013 IEEE International Conference on (pp. 8604-8608). IEEE.

\bibitem{mfcc} Zheng, F., Zhang, G., Song, Z. ``Comparison of different implementations of MFCC.'' Journal of Computer Science and Technology 16.6 (2001): 582-589.

\bibitem{wu2016} Wu, K.,  Zhang, D., Lu, G. ``iPEEH: Improving pitch estimation by enhancing harmonics.'' Expert Systems with Applications 64 (2016): 317-329.

\bibitem{han2014} Han, K., DeLiang, W. ``Neural network based pitch tracking in very noisy speech.'' IEEE/ACM Transactions on Audio, Speech and Language Processing (TASLP) 22.12 (2014): 2158-2168.

\bibitem{seltzer2013investigation} Seltzer, M.L., Yu, D., Wang, Y. ``An investigation of deep neural networks for noise robust speech recognition.'' Acoustics, Speech and Signal Processing (ICASSP), 2013 IEEE International Conference on. IEEE, 2013.

\bibitem{abdel2012applying} Ossama, A. et al. ``Applying convolutional neural networks concepts to hybrid NN-HMM model for speech recognition.'' Acoustics, Speech and Signal Processing (ICASSP), 2012 IEEE International Conference on. IEEE, 2012.

\bibitem{huang2013audio} Jing, H., Kingsbury, B. ``Audio-visual deep learning for noise robust speech recognition.'' Acoustics, Speech and Signal Processing (ICASSP), 2013 IEEE International Conference on. IEEE, 2013.

\bibitem{kumar2016} Kumar, A., Florencio, D. ``Speech Enhancement In Multiple-Noise Conditions using Deep Neural Networks.'' arXiv preprint arXiv:1605.02427 (2016).

\bibitem{hinton2012deep} Hinton, G., et al. ``Deep neural networks for acoustic modeling in speech recognition: The shared views of four research groups.'' Signal Processing Magazine, IEEE 29.6 (2012): 82-97.

\bibitem{coto2017} Coto-Jim\'enez, M., Goddard-Close, Mart\'inez-Licona, F.M. ``Improving Automatic Speech Recognition Containing Additive Noise Using Deep Denoising Autoencoders of LSTM Networks.'' In: Ronzhin A., Potapova R., Németh G. (eds) Speech and Computer. SPECOM 2016. \emph{Lecture Notes in Computer Science}, vol. 9811. Springer, Cham. doi: 10.1007/978-3-319-43958-7\_42

\bibitem{feng2014speech} Feng, X., Zhang, Y., Glass, J. ``Speech feature denoising and dereverberation via deep autoencoders for noisy reverberant speech recognition.'' Acoustics, Speech and Signal Processing (ICASSP), 2014 IEEE International Conference on. IEEE, 2014.

\bibitem{ishii2013} Ishii, T., Komiyama, H., Shinozaki, T., Horiuchi, Y. and Kuroiwa, S., 2013, August. Reverberant speech recognition based on denoising autoencoder. In INTERSPEECH (pp. 3512-3516).

\bibitem{nakashika-2013} Nakashika, T., Takashima, T., Takiguchi, T., Ariki, Y. Voice conversion in high-order eigen space using deep belief nets. {\em Proceedings of Interspeech} {\bf 2013}.

\bibitem{chen-2015} Chen, L.H. et al. A deep generative architecture for postfiltering in statistical parametric speech synthesis. {\em IEEE/ACM Transactions on Audio, Speech and Language Processing (TASLP)} {\bf 2015} {\em 23.11}, 2003--2014. doi: 10.1109/TASLP.2015.2461448.

\bibitem{coto-2017} Coto-Jim\'enez, M., Goddard-Close, J. LSTM Deep Neural Networks Postfiltering for Enhancing Synthetic Voices. {\em International Journal of Pattern Recognition and Artificial Intelligence} {\bf 2018} {\em 32.3} (2018). Doi: https://doi.org/10.1142/S021800141860008X.


\bibitem{coto-2016c} Coto-Jim\'enez, M., Goddard-Close, J. LSTM Deep Neural Networks Postfiltering for Improving the Quality of Synthetic Voices. {\em Mexican Conference on Pattern Recognition} {\bf 2016}.

\bibitem{liu2017} Liu, B. et al. ``A novel pitch extraction based on jointly trained deep BLSTM Recurrent Neural Networks with bottleneck features.'' Acoustics, Speech and Signal Processing (ICASSP), 2017 IEEE International Conference on. IEEE, 2017.

\bibitem{dahl2012} Dahl, G. et al. ``Context-dependent pre-trained deep neural networks for large-vocabulary speech recognition.'' IEEE Transactions on audio, speech, and language processing 20.1 (2012): 30-42.

\bibitem{erhan2010} Erhan, D. et al. ``Why does unsupervised pre-training help deep learning?.'' Journal of Machine Learning Research 11.Feb (2010): 625-660.

\bibitem{vanden2014} Van Den Oord, A., Dieleman, S. Schrauwen, B. ``Transfer learning by supervised pre-training for audio-based music classification.'' Conference of the International Society for Music Information Retrieval (ISMIR 2014). 2014.

\bibitem{donahue2014} Donahu, J. et al. ``Decaf: A deep convolutional activation feature for generic visual recognition.'' International conference on machine learning. 2014.

\bibitem{vesely2013} Vesely, K. Hannemann, H., Burget, L. ``Semi-supervised training of deep neural networks.'' Automatic Speech Recognition and Understanding (ASRU), 2013 IEEE Workshop on. IEEE, 2013.

\bibitem{ocho} Fan, Y., Qian, Y.. Xie, F.L., Soong, F.K. ``TTS synthesis with bidirectional LSTM based recurrent neural networks.'' \emph{Proceedings of Interspeech}, 2014.

\bibitem{trece} Sepp, H., Schmidhuber, J. ``Long short-term memory.'' \emph{Neural computation}, vol. 9, no. 8, pp. 1735-1780, 1997.

\bibitem{diez} Graves, A., Fern\'andez, S., Schmidhuber, J. ``Bidirectional LSTM networks for improved phoneme classification and recognition.'' {\it International Conference on Artificial Neural Networks}. Springer Berlin Heidelberg, 2005.

\bibitem{once} A. Graves, N. Jaitly, and A. Mohamed. ``Hybrid speech recognition with deep bidirectional LSTM.'' {\it IEEE Workshop on Automatic Speech Recognition and Understanding (ASRU)}, 2013.

\bibitem{gers2002} Gers, F. A., Schraudolph, N. N., \& Schmidhuber, J. (2002). Learning precise timing with LSTM recurrent networks. Journal of machine learning research, 3(Aug), 115-143.

\bibitem{vincent2010stacked} Pascal, V. et al. "Stacked denoising autoencoders: Learning useful representations in a deep network with a local denoising criterion." The Journal of Machine Learning Research 11 (2010): 3371-3408.

\bibitem{tres} Chen, L.H., Raitio, T., Valentini-Botinhao, C., Ling, Z.H., Yamagishi, J. ``A deep generative architecture for postfiltering in statistical parametric speech synthesis.'' \emph{IEEE/ACM Transactions on Audio, Speech and Language Processing (TASLP)}, vol.23, no. 11, pp. 2003-2014, 2015. doi: 10.1109/TASLP.2015.2461448.

\bibitem{baek2003} Baek, J., Cho, S. ``Bankruptcy prediction for credit risk using an auto-associative neural network in Korean firms.'' Computational Intelligence for Financial Engineering, 2003. Proceedings. 2003 IEEE International Conference on. IEEE, 2003.

\bibitem{cmu} Kominek, J., Black, AW (2004): The CMU Arctic speech databases. Fifth ISCA Workshop on Speech Synthesis.

\bibitem{erro2} Erro, D., Sainz, I., Saratxaga, I.,Navas, E., Hern\'aez, I. ``MFCC+F0 extraction and waveform reconstruction using HNM: preliminary results in an HMM-based synthesizer'', VI Jornadas en Tecnologia del Habla \& II Iberian SLTech (FALA), pp. 29-32, 2010.





























\end{thebibliography}
%

\end{document}